\documentclass[%
superscriptaddress,
preprint,
 amsmath,amssymb,
 aps,
prb,
]{revtex4-1}

\pdfoutput=1
\usepackage{graphicx}
\usepackage{dcolumn}
\usepackage{bm}
\usepackage{setspace}
\usepackage{hyperref}
\usepackage{placeins}
\usepackage{subfig}
\usepackage{balance}  

\begin{document}

\newcommand{\bo}{\boldsymbol}
\newcommand{\boq}{\mathbf{q}}
\newcommand{\bok}{\mathbf{k}}
\newcommand{\bor}{\mathbf{r}}
\newcommand{\boG}{\mathbf{G}}
\newcommand{\boR}{\mathbf{R}}
\newcommand\2{$_2$}

\newcommand\theosmarvel{Theory and Simulation of Materials (THEOS), and National Centre for Computational Design and Discovery of Novel Materials (MARVEL), \'Ecole Polytechnique F\'ed\'erale de Lausanne, CH-1015 Lausanne, Switzerland}

\title{Prediction of phonon-mediated superconductivity  with high critical temperature in the two-dimensional topological semimetal W$_2$N$_3$}

\author{Davide Campi}
\affiliation{\theosmarvel}
\author{Simran Kumari}
\affiliation{\theosmarvel}
\author{Nicola Marzari}
\affiliation{\theosmarvel}

\date{\today}



\begin{abstract}
Two-dimensional superconductors attract  great interest both for their  fundamental physics and for their potential applications, especially in the rapidly growing field of quantum computing. Despite intense theoretical and experimental efforts, materials with a reasonably high transition temperature are still rare.  Even more rare are those that combine superconductivity with a non-trivial band topology, to potentially host exotic states of matter such as Majorana fermions. Here, we predict a remarkably high superconducting critical temperature of 21-28 K in the easily exfoliable, topologically non-trivial 2D semimetal W$_2$N$_3$. By studying its  electronic and superconducting properties as a function of doping and strain, we find large changes in the electron-phonon interactions that make this material a unique platform to study different coupling regimes and test the limits of  current theories of superconductivity. Last, we discuss the possibility of tuning the material to achieve coexistence of superconductivity and topologically non-trivial edge states.

\end{abstract}

\maketitle

\subsection{Introduction}
In recent years, superconductivity in two-dimensional (2D) systems has attracted  great and ever-increasing interest, thanks both to its relevance for fundamental physics understanding and its potential technological applications for emergent nanoscale devices such as quantum interferometers, superconducting transistors, and superconducting qubits\cite{DeFranceschi2010,Huefner2009,Delahaye2003,Romans2010,Liu2019}.

After the ground-breaking work of Zhang et al., demonstrating in 2010 superconductivity up to 1.8 K in a single Pb layer on Si(111)\cite{Zhang2010}, the field of highly crystalline 2D superconductors has developed rapidly, both theoretically and experimentally, moving from single-layer metallic films deposited by molecular-beam epitaxy to intrinsically 2D monolayers derived from weakly-bonded layered materials\cite{Saito2016,Brun2016,Saito2016iop}. Electrostatically-doped or alkali-decorated graphene was predicted soon after  to undergo a superconducting transition\cite{Savini2010,Profeta2012,Margine2014,Zheng2016}. Experimental evidence of such transition has been reported for K-intercalated few-layer graphene at 4.5 K\cite{Xue2012}, for  Ca-intercalated epitaxial graphene at 7 K \cite{Li2013}, as well as for Li-intercalated graphene \cite{Ludbrook2015,Tiwari2017}.

More recently, superconductivity has been observed in gated few and single layers of transition-metal dichalcogenides, with a critical temperature of 7.2 K for NbS$_2$\cite{Frindt1972,Ugeda2015}, around 5.3 K for NbSe$_2$\cite{Frindt1972,Ugeda2015,Tsen2016,Xi2016}, 3 K for TiSe$_2$\cite{Li2016} and between 7 and 12 K for MoS$_2$ \cite{Ye2012,Lu2015,Costanzo2016,Fu2017}.
A surprising enhancement of the superconducting transition temperature upon dimensional reduction has also been reported for TaS$_2$ from 0.5 to 2.2 K \cite{NavarroMoratalla2016}. 
 
Various theoretical efforts have been made to find 2D superconductors with higher T$_c$, including several doped 2D materials \cite{Shao2014,Huang2015,Huang2016,Sanna2016,Lugovskoi2019} and intrinsic 2D metals \cite{Lei2017,Zhang2017}, with predicted transition temperatures ranging from 10 to 20 K. 2D boron allotropes\cite{Mannix2015,Feng2016},recently realized, have attracted considerable interest due to several theoretical predictions of a superconducting transition at a temperature above liquid hydrogen \cite{Zhao2016,Zhao2016b,Xiao2016,Penev2016,Gao2017,Li2018}. However, unlike most of the aforementioned materials, 2D boron allotropes cannot be obtained by exfoliation from van der Waals-bonded 3D parents, but they have to be grown directly on a metal substrate. This results in relatively strong interactions with the substrate, which are believed to suppress the superconducting critical temperature down to 2 K \cite{Cheng2017}.

Independently from the quest for higher T$_c$, the search for materials combining nontrivial topological properties with superconductivity has also been the subject of intense investigations driven by the quest for exotic states of matter, such as Majorana fermions, that can arise from the interaction between topological edge states and the superconducting phase\cite{Xu2014,Sarma2015,Sato2017}. Recently, a great interest was aroused by the experimental confirmation of superconductivity below 1 K in the electrostatically-doped 2D topological insulator WTe2 \cite{Sajadi2018,Fatemi2018}.

In this paper we find, by mean of first-principles calculations,  intrinsic superconductivity above the temperature of liquid hydrogen in monolayer W$_2$N$_3$. We discuss how such system could be potentially tuned to exploit the coexistence of superconductivity and non-trivial band topology, and highlight a very strong sensitivity of the electron-phonon coupling with strain and doping that makes W$_2$N$_3$ a promising playground to test different coupling regimes.

\subsection{Structural, electronic and topological properties}

Two-dimensional W$_2$N$_3$ has been recently identified in the first-principle calculations of Mounet et al.\cite{Mounet2018} as easily exfoliable from the layered hexagonal-W$_2$N$_3$  bulk; this later was first reported experimentally by Wang et al. in 2012\cite{Wang2012}. The binding energy of monolayer W$_2$N$_3$ is 26.3 meV/\AA$^{2}$, very close to the values computed for the most common transition-metals dichalcogensides, making it an ideal candidate for mechanical exfoliation. 

Bulk hexagonal-W$_2$N$_3$ was believed to be metastable at ambient pressure\cite{Mehl2015}, and its existence in an ordered layered structure has been recently questioned in favor of a disordered W$_{2.25}$N$_3$ structure with partial occupations of the W sites\cite{Kawamura2018}. However, indisputable experimental evidence of an ordered layered structure has recently been provided, as well as the possibility of exfoliation down to the monolayer\cite{Jin2019,Huang2020}. 

\begin{figure}[h!]
\includegraphics[width=0.45\textwidth]{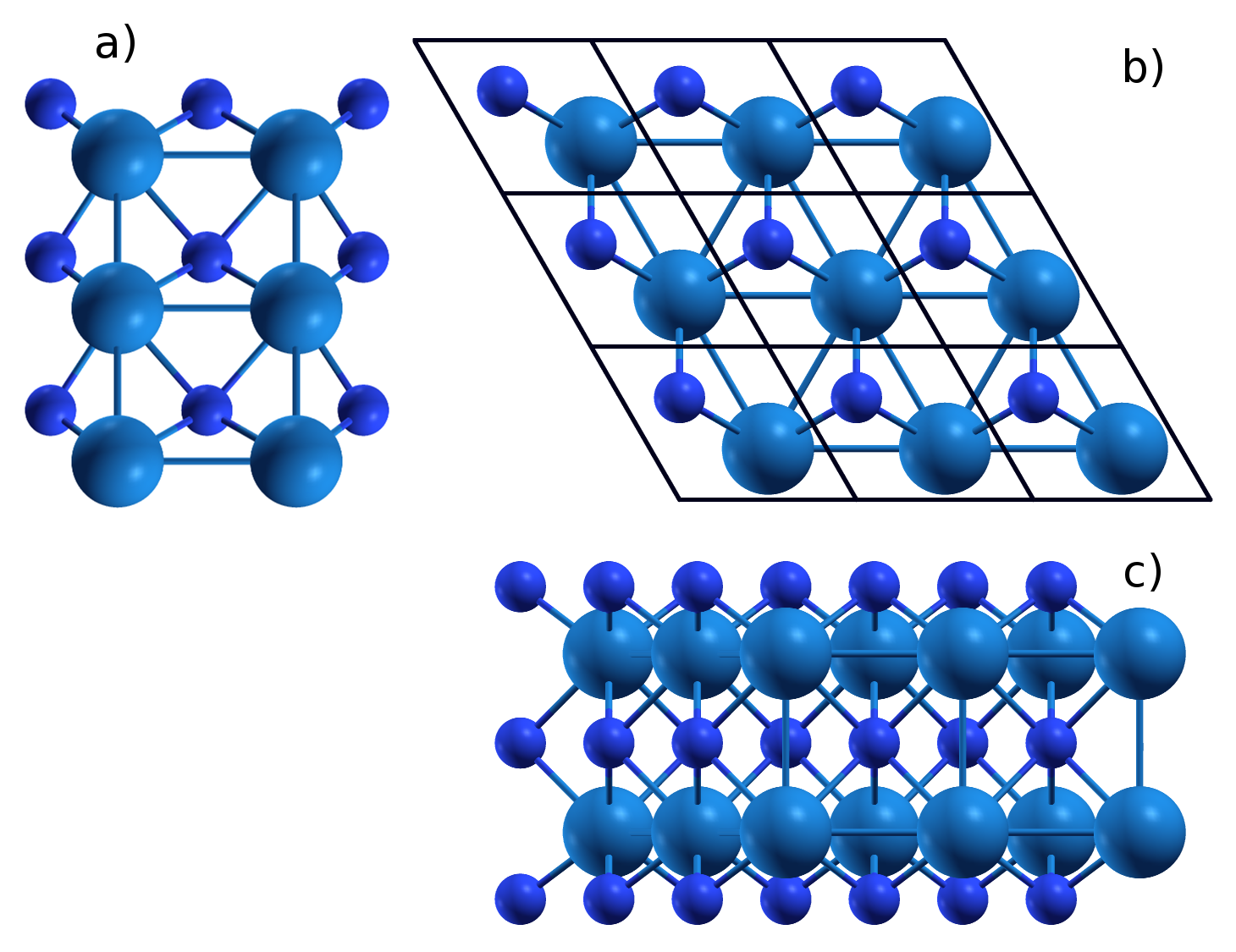}
\caption{Crystal structure of W$_2$N$_3$ as seen from the x axis (a), the y axis (c) and the z axis (b). The primitive cell is also shown.}
\label{fig:structure}
\end{figure} 

The crystal structure of 2D W$_2$N$_3$ is schematically represented in Fig.\ref{fig:structure}. The material is characterized by an hexagonal unit cell with a P-6m2 space group where the two six-coordinated W atoms occupies the 3c(0, 1/2, 1/2) Wyckoff sites, two three-coordinated N anions occupies the 3d(1/2, 0, 0) sites while a central five-coordinated N occupy the 1b(1/2, 1/2, 1/2) site. 

We optimized the 2D structure using density-functional theory in the GGA-PBE approximation\cite{pbe} with plane waves and norm-conserving pseudopotentials \cite{Pseudodojo},
as implemented in the Quantum-ESPRESSO distribution\cite{QMEspresso}. The plane wave kinetic energy cutoff is 80 Ry and the structural optimization is performed until forces on atoms are less than 10 meV/\AA$^{2}$.
The Brillouin zone (BZ) has been sampled with a uniform unshifted 16$\times$16$\times$1 \textbf{k}-point mesh\cite{mp} and a Gaussian smearing of 0.01 Ry has been adopted, to deal with the metallic character of the material.   40 \AA\ of vacuum separate periodic layers to avoid spurious interactions. Calculations have been carried out both with and without spin-orbit coupling (SOC). The optimized theoretical equilibrium in-plane lattice parameter is 2.912 \AA\ (without SOC)  and 2.898 \AA\ (including SOC), both very close to the measured bulk value of 2.890 \AA\cite{Wang2012}, hinting again at the weakness of the interlayer interactions.


\begin{figure}[h!]
\includegraphics[width=0.47\textwidth]{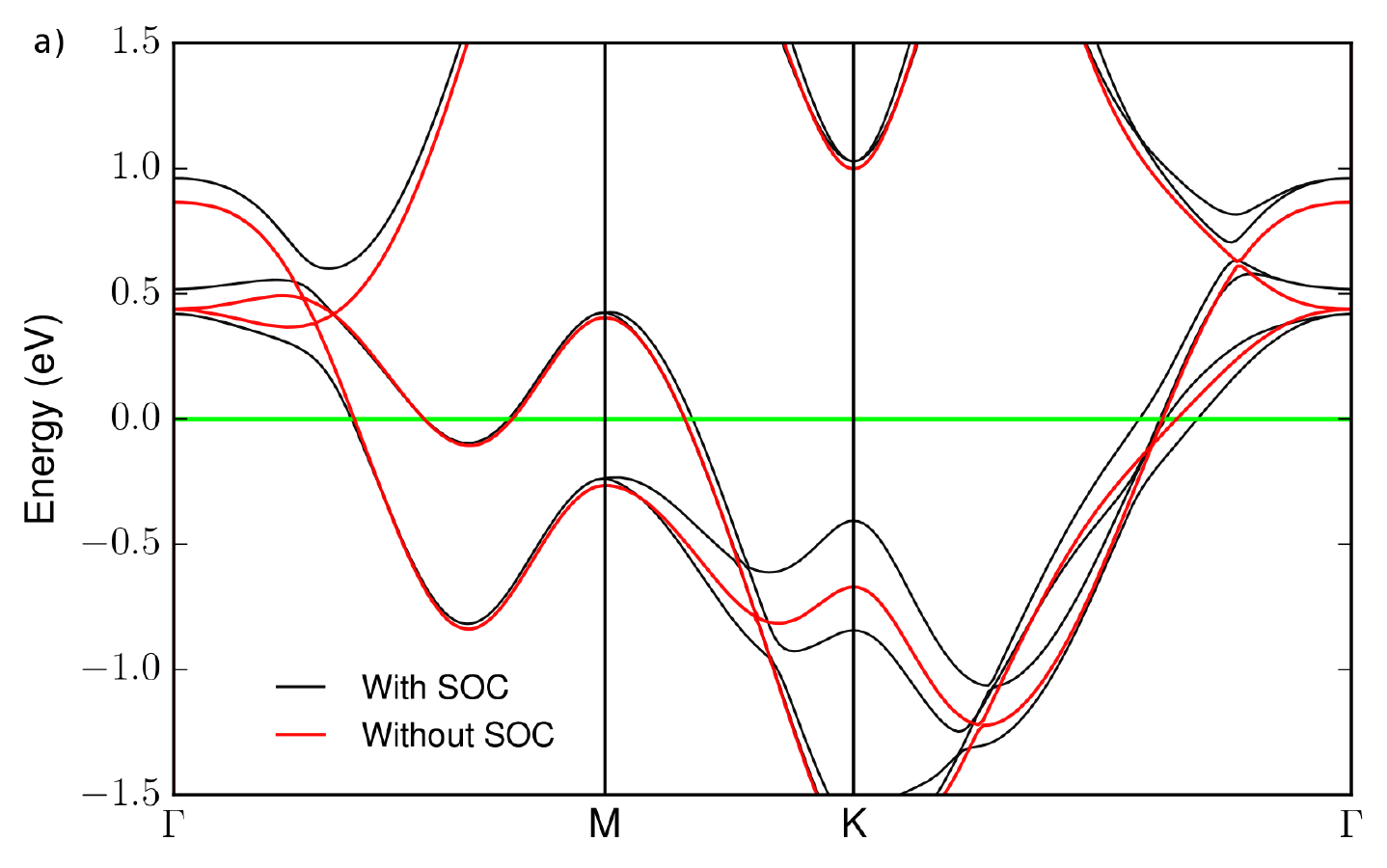}
\includegraphics[width=0.45\textwidth]{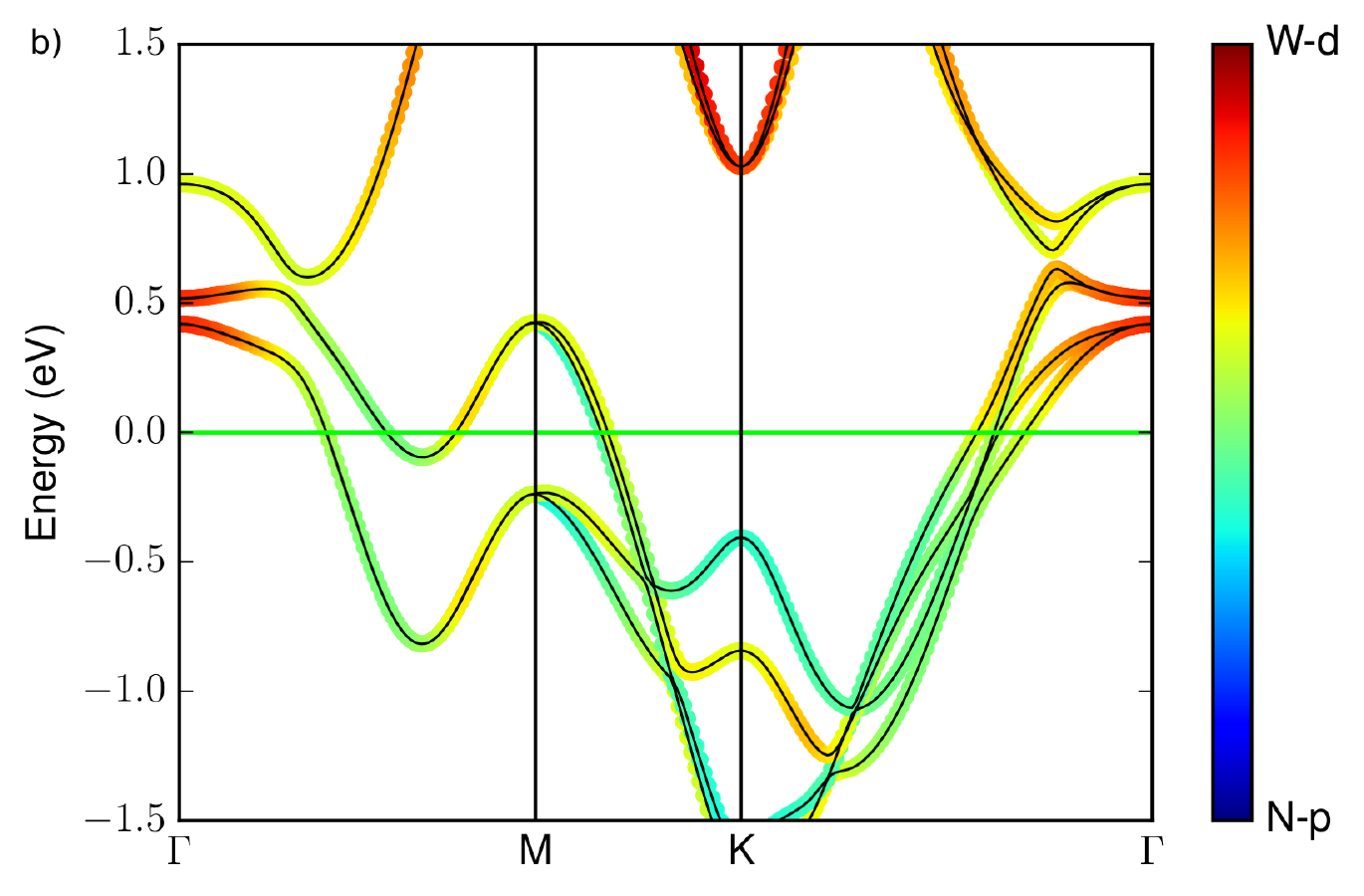}
\caption{a) Electronic band structure along a high-symmetry path with (black) or without (red) spin-orbit coupling. Note the non-trivial  opening of local gaps at around 0.6 eV above the Fermi energy (solid green line). b) Band structure showing the major orbital contributions. Bands arise from a strong hybridization of d states from the W atoms and p states from the outermost N atoms. The color code is proportional to the prevalent character of the band (W d in red, N p in blue).}
\label{fig:bands}
\end{figure} 

We show in Fig.\ref{fig:bands} the calculated band structure along a high-symmetry path, respectively with and without SOC. The major contributions near the Fermi level come from the W d bands, that are strongly hybridized with the outermost p-states of the N atoms in the (3d) position. (see Fig.\ref{fig:bands} b)). Spin-orbit coupling splits the degeneracy of the 2 states lying 0.5-0.6 eV above the Fermi energy and, noticeably, it opens a gap in the nodal line around $\Gamma$ giving origin to a non-trivial band gap of 0.1 eV  approximately 0.6 eV above the Fermi energy. The topologically non-trivial nature of this material, first proposed by Wang et al.\cite{Wang2019} on the basis of symmetry considerations, is related to existence of a mirror Chern number as discussed in more details in the supplementary information (S.I.). 

  
\begin{figure}[h!]
\includegraphics[width=0.5\textwidth]{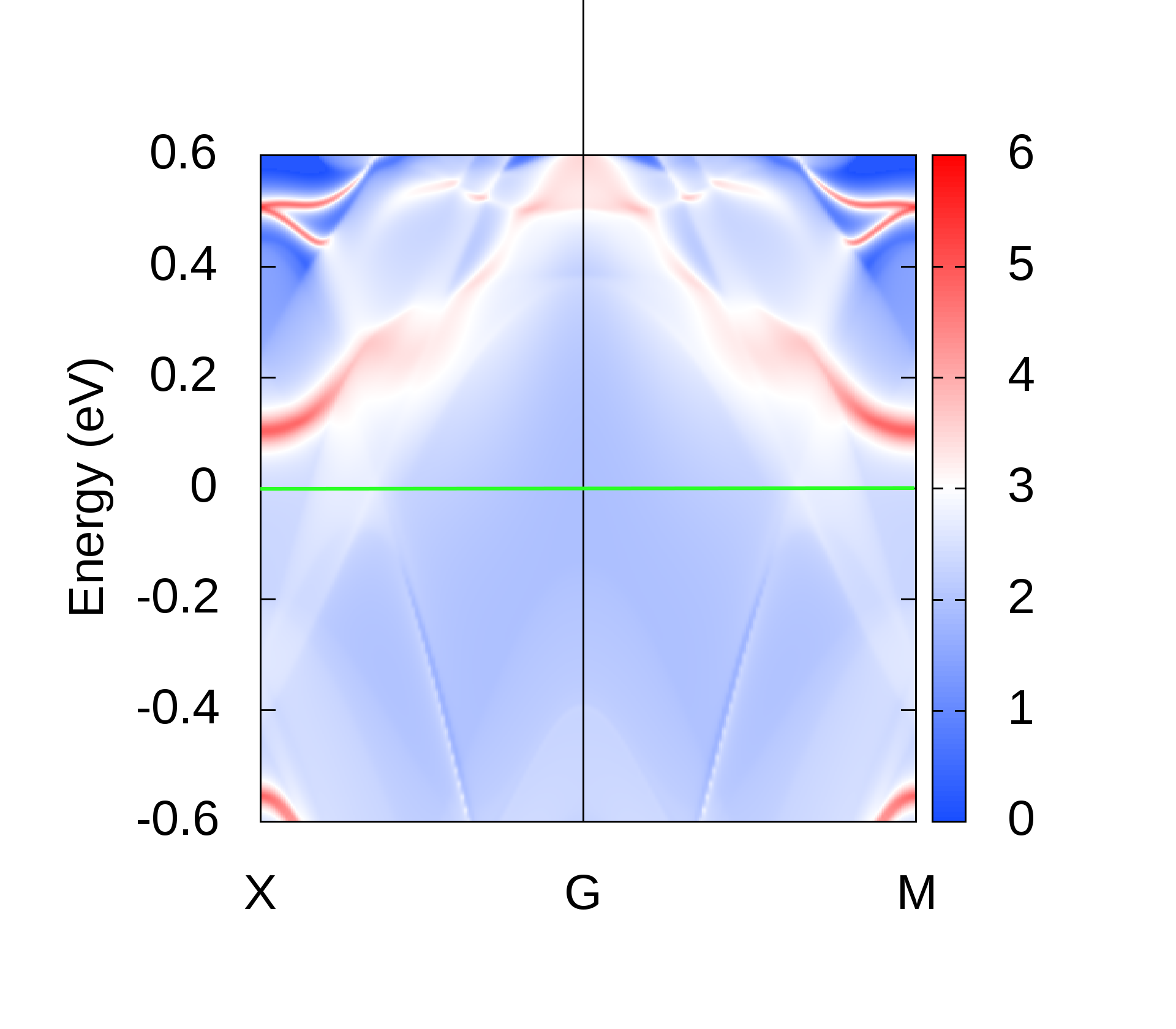}
\caption{Projected edge states  for a (1, 0) W$_2$N$_3$ nanowire studied using a repeated non-primitive 4-atoms rectangular unit cell. The topologically protected states can be identified closing the gap at the zone border 0.5 eV above the Fermi level. }
\label{fig:topology}
\end{figure} 

To highlight the topological nature of monolayer W$_2$N$_3$ we report in Fig.\ref{fig:topology} the projected edge states for a (1,0) nanowire. The calculation has been performed for a non primitive, 4-atom rectangular unit cell, using WannierTools\cite{Wu2017} with a  tight-binding model based on maximally localized Wannier functions, computed by Wannier90\cite{Mostofi2014}. At the Kohn-Sham DFT level, the unpopulated topological edge states can be found between 0.4 and 0.6 eV above the Fermi energy connecting at the high-symmetry points at the zone border. These states, once occupied, might interact with the superconducting state, possibly giving rise to more exotic features.

\subsection{Intrinsic electron-phonon coupling}
To study the superconducting properties of 2D W$_2$N$_3$ we start by computing the phonon dispersions with density-functional perturbation theory\cite{Baroni} on a 8$\times$8 mesh of \textbf{q}-points in the Brillouin zone. For the electron-phonon coupling coefficients we employ two independent methods: 1) a direct calculation on a dense electronic \textbf{k}-mesh followed by a linear interpolation of linewidths on a dense phonon grid directly in Quantum ESPRESSO, 2) a Wannier interpolation as implemented in the EPW code\cite{Giustino2017,epw2010,epw2016}. 
In the first case we use a 192$\times$192 \textbf{k}-point electronic grid and interpolate on a 96$\times$96 \textbf{q}-point  phonon grid, estimating the transition temperature with a standard McMillan-Allen-Dynes formula\cite{McMillan1968,Allen1975}. With EPW, we solve instead the Migdal-Eliashberg equations\cite{Margine2013} both in the isotropic and anisotrpic approximations to obtain the superconducting gap and its temperature evolution.  For the isotropic Eliashberg equations we use 192$\times$192 and 96$\times$96 \textbf{k}- and \textbf{q}-point grids respectively, while for the more demanding anisotropic case we halve both grids. In all cases 1 eV cutoff for the Matsubara frequency is chosen to be five times the largest phonon frequency, and the Dirac delta functions are replaced by Lorentzians of widths 50 and 0.5 meV for electrons and phonons, respectively. 
A value of 0.16 is used for the screened Coulomb parameter $\mu^{\ast}$ (this semiempirical screened Coulomb parameter is usually taken between 0.05 and 0.2 for 2D materials with a mean value of 0.1\cite{Profeta2012,Margine2014,Zhao2016,Zhao2016b,Xiao2016,Penev2016,Gao2017,Li2018}). A lower value of 
 $\mu^{\ast}$ typically implies an increase in the critical transition temperature and we verified the sensitivity of our results showing an increase of at most 10\% in T$_c$ when $\mu^{\ast}$=0.1 is chosen for the undoped case.


\begin{figure}[h!]
\includegraphics[width=0.55\textwidth]{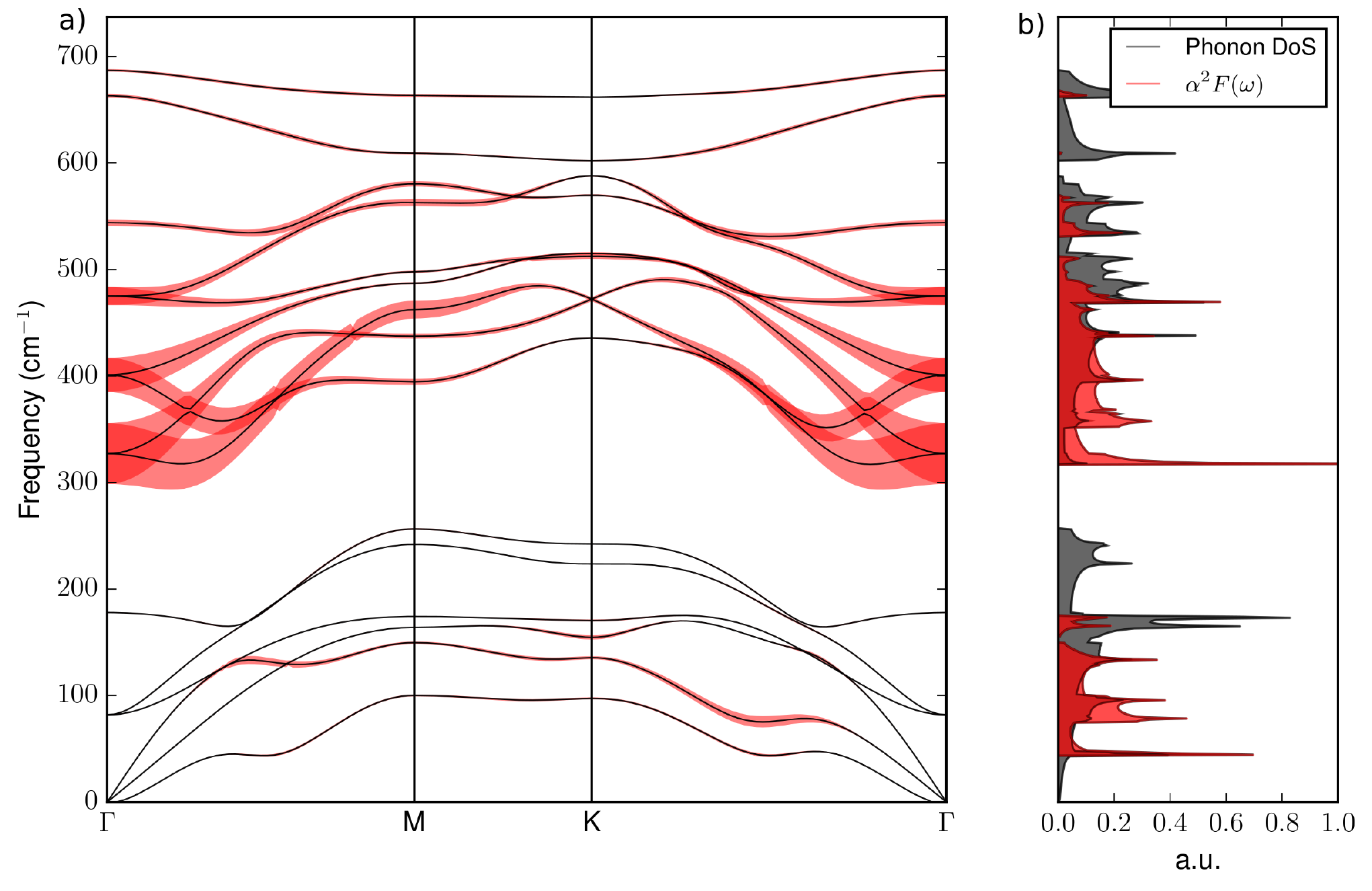}
\includegraphics[width=0.55\textwidth]{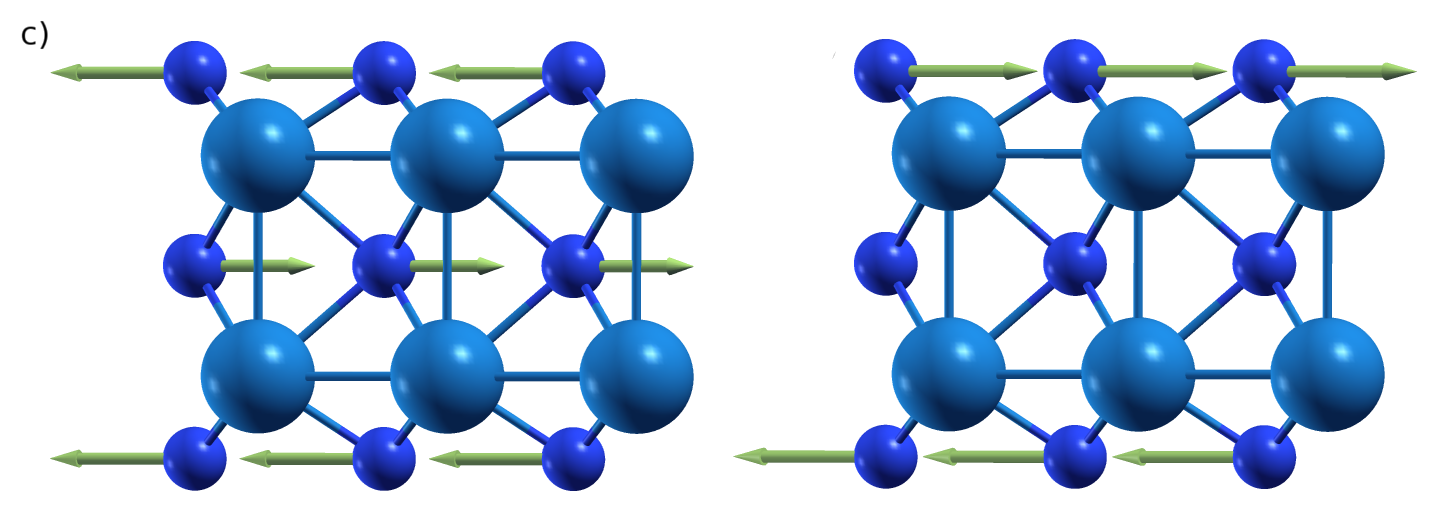}
\caption{a) Phonon dispersions phonon linewidths (magnified by a factor 10) along a standard high-symmetry path. b) Eliashberg function and phonon density of states as a function of the phonon energy. The most important contributions to the electron-phonon interaction come from an optical mode modulating in-plane the outermost N-W bond, and from the two lowest acoustic modes (longitudinal and shear horizontal) near the middle of the $\Gamma$-M and $\Gamma-K$ paths. c) Eigenvectors of the two phonon modes at 330 and 400 cm$^{-1}$ mostly responsible for the electron-phonon coupling in the optical region.}
\label{fig:phonons} 
\end{figure} 

The phonon dispersion relations and the respective electron-phonon linewidths are reported in Fig.\ref{fig:phonons} a), together with the phonon density of states and the Eliashberg function as a function of energy. We can recognize two main contributions to the electron-phonon interactions; the first comes from the longitudinal and shear horizontal acoustic modes near the center of the Brillouin zone, where the coupling is strong enough to renormalize the phonon frequencies and induce an anomaly in the dispersions, signaling an incipient charge-density-wave instability. The second one comes from the optical phonon modes at the zone center, with frequencies of 330 and 400 cm$^{-1}$, showing respectively the largest and the second largest overall linewidths. Both modes are characterized by a modulation of the in-plane bond between the outermost N atoms and the W ones. 

\begin{figure}[h!]
\includegraphics[width=0.4\textwidth]{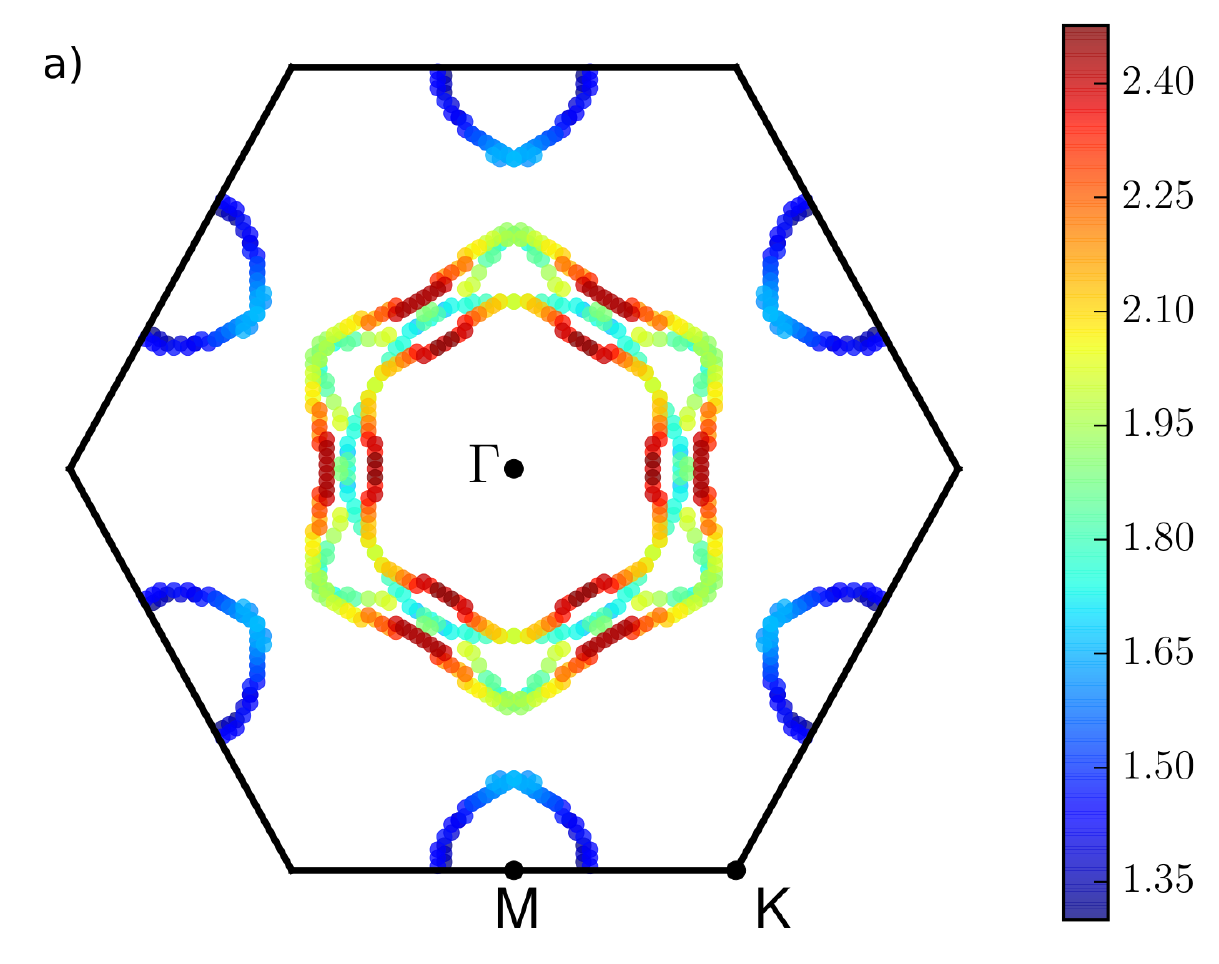}
\includegraphics[width=0.4\textwidth]{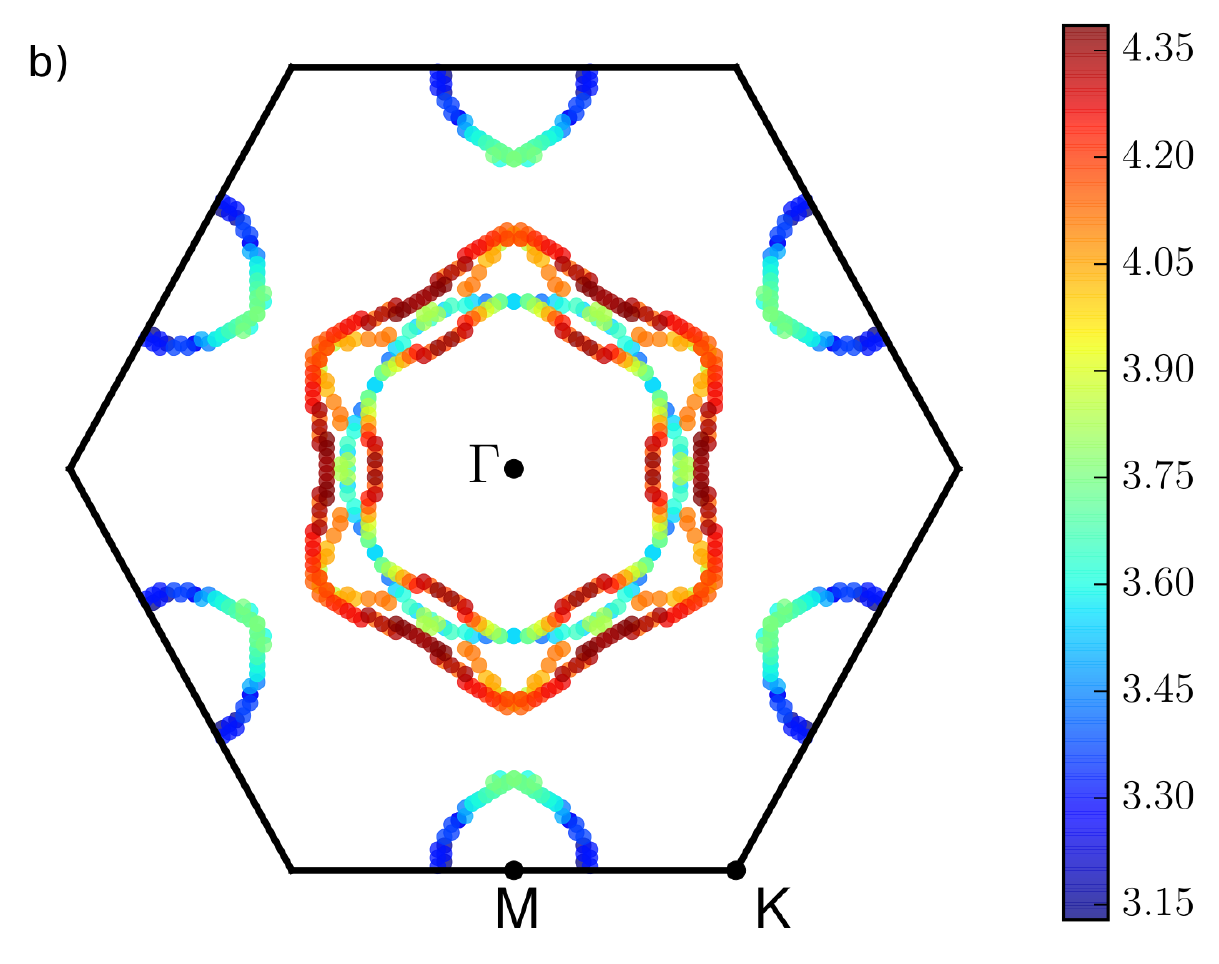}
\caption{a) Momentum-dependent electron electron-phonon coupling constant ($\lambda_{\bf{k}}$) over the full Brillouin zone and b) momentum-dependent superconducting band gap at 10 K projected on the Fermi lines. Both the electron-phonon coupling and the gap are sensibly anisotropic with the largest components along the $\Gamma$-K direction in the two central rings around }  
\label{fig:fermi2} 
\end{figure}

\begin{figure}[h!]
\includegraphics[width=0.45\textwidth]{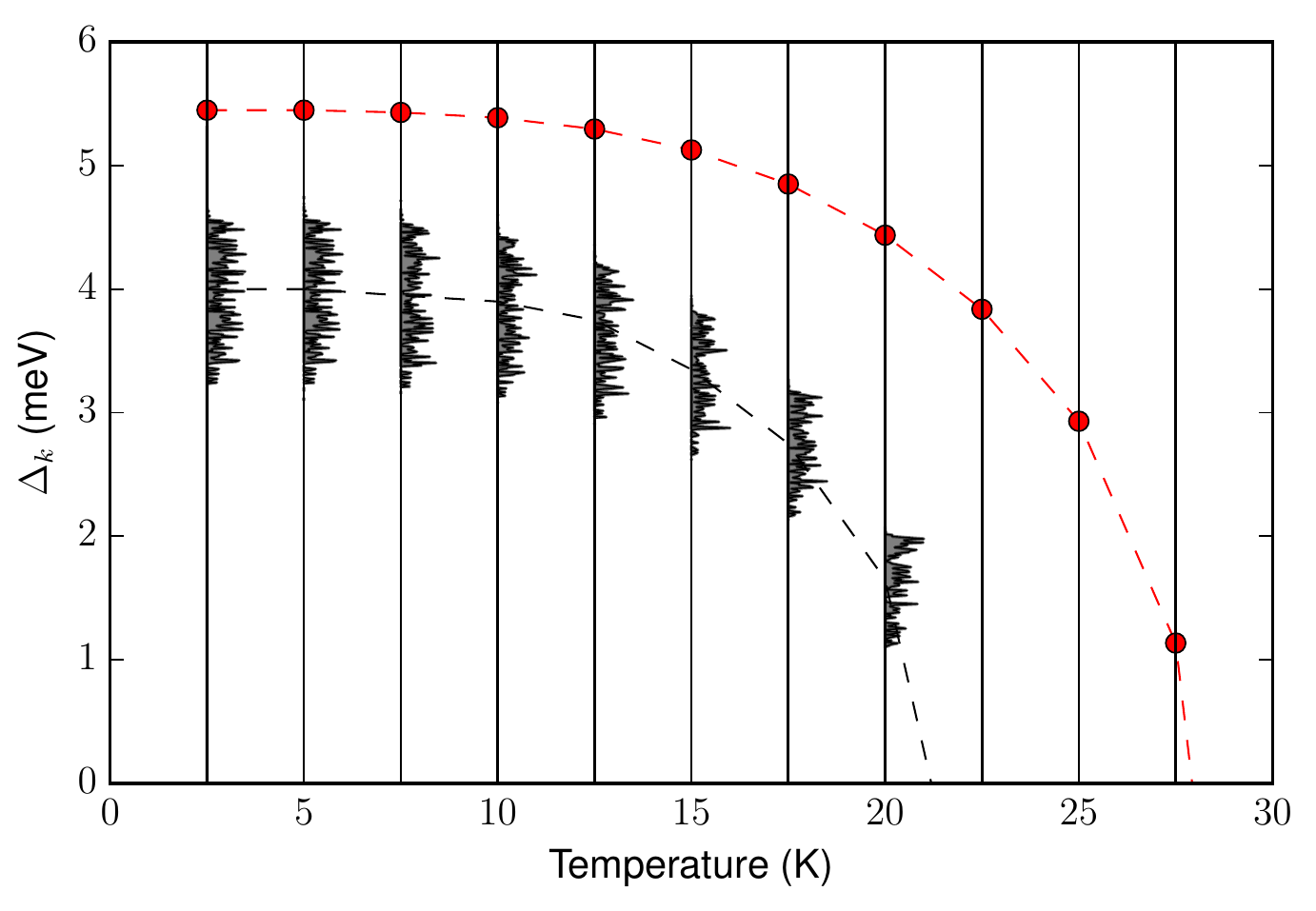}
\caption{Evolution of the superconducting gap $\Delta_{\textbf{k}}$ as a function of temperature, computed by solving the  Migdal-Eliashberg equations in the isotropic approximation (red dots and dashed line interpolation) and with a fully anisotropic solution where the grey histograms show the magnitude distribution of the momentum-dependent superconducting gap and the gray dashed line interpolates the centers of the histograms.}
\label{fig:scgap}
\end{figure}

We report in figs. \ref{fig:fermi2}  the momentum-dependent electron-phonon couplings $\lambda_{\bf{k}}$ and the superconducting gap $\Delta_{\textbf{k}}$ at 10 K. Both quantities display a similar anisotropy, with their maximum along the $\Gamma$-K direction, where the two concentric outermost and innermost hexagonally-warped Fermi lines find  preferential nesting around $\Gamma$. The coupling is nearly 20\% weaker in the $\Gamma$-M direction and more than 25\% weaker in the other two bands. A more distinct difference in the magnitude of the coupling and the superconducting gap can instead be noticed with the pockets around the M point, characterized by a much weaker coupling (almost 50\% weaker) hinting towards a two-gap structure, even if this structure is not particularly evident in the distribution of the momentum dependent superconducting gap (see Fig. \ref{fig:scgap} due to a partial overlapping in the magnitude distribution of the two gaps.

Figure \ref{fig:scgap} shows the evolution of the superconducting gap as a function of temperature,  computed by solving the  Migdal-Eliashberg equations in both the isotropic and in the fully anisotropic approximations. The transition temperature is identified as the lowest temperature at which a vanishing of the superconducting gap is observed. The superconducting transition temperature computed within the isotropic approximation is higher (28 K) than the one computed with the fully anisotropic one (21 K); this latter in turn agrees well with the value obtained from the simpler Allen-Dynes estimate (20.4 K), using the same code and dense electron and phonon grids. The value is also very close with the one obtained from brute force integration and linear interpolation with Quantum ESPRESSO (19.4 K). 
 We can conclude that even the lowest estimate presented here would be record-high for a 2D material, making intrinsic W$_2$N$_3$ a very promising solution for high-temperature superconductivity in 2D.

\subsection{Strain and doping}

At variance with bulk systems, 2D materials offer a greater possibility of manipulation by doping or by applying an external strain. Both those approaches have been used or proposed to tune the superconducting properties of mono- and bi-layers, either by electrostatic gating \cite{Ugeda2015,Costanzo2016,NavarroMoratalla2016}, intercalation\cite{Profeta2012,Huang2015,Zheng2016,Sanna2016} or strain \cite{Zhang2017,Li2018}. In this Section we study how strain and doping can be used to tune the electron-phonon interactions in W$_2$N$_3$ to either reach regimes of extreme coupling on the edge of a charge-density-wave instability, or possibly take advantage of W$_2$N$_3$ topologically non-trivial nature.

\begin{figure}[h!]
\includegraphics[width=0.5\textwidth]{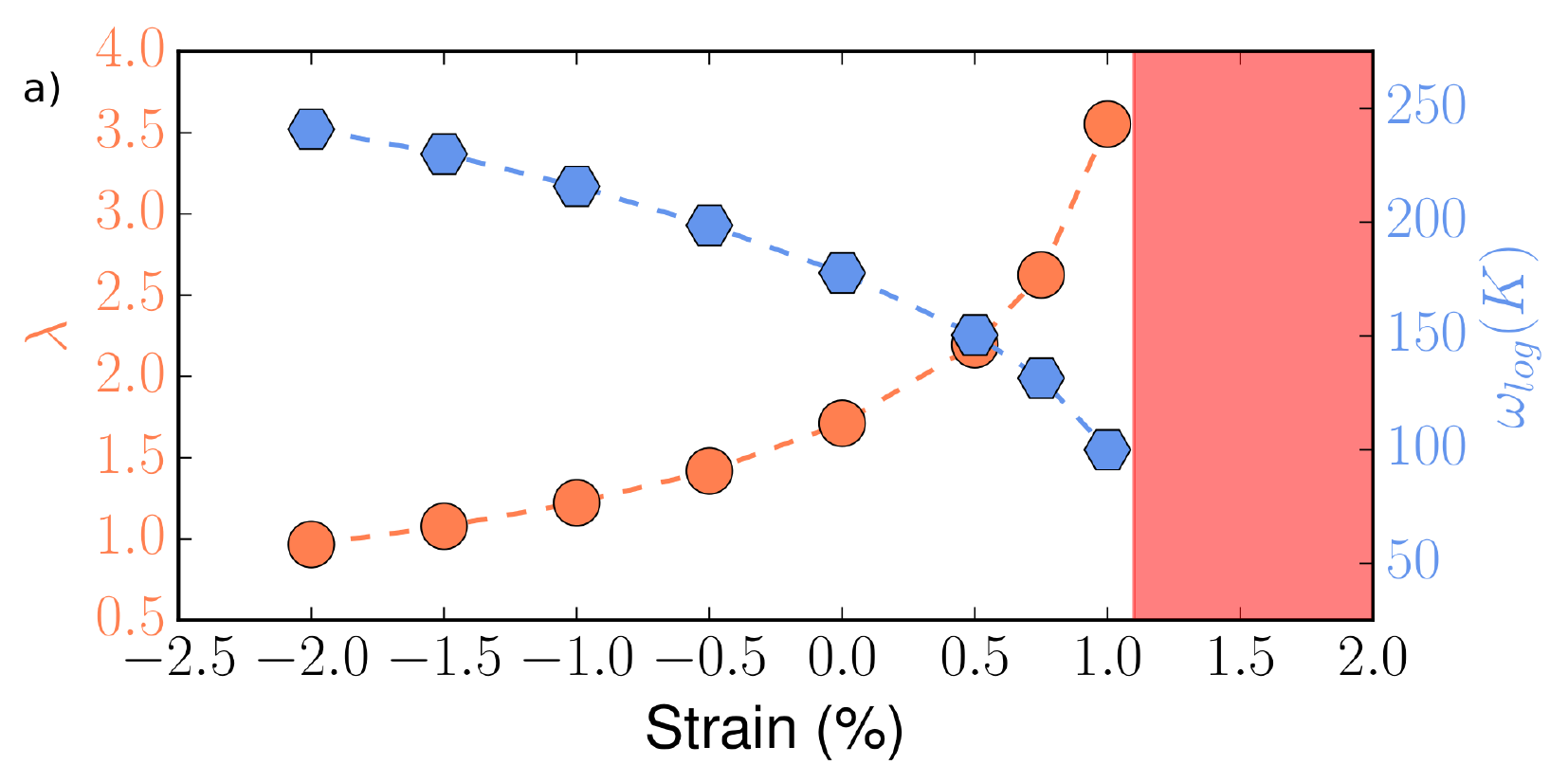}
\includegraphics[width=0.45\textwidth]{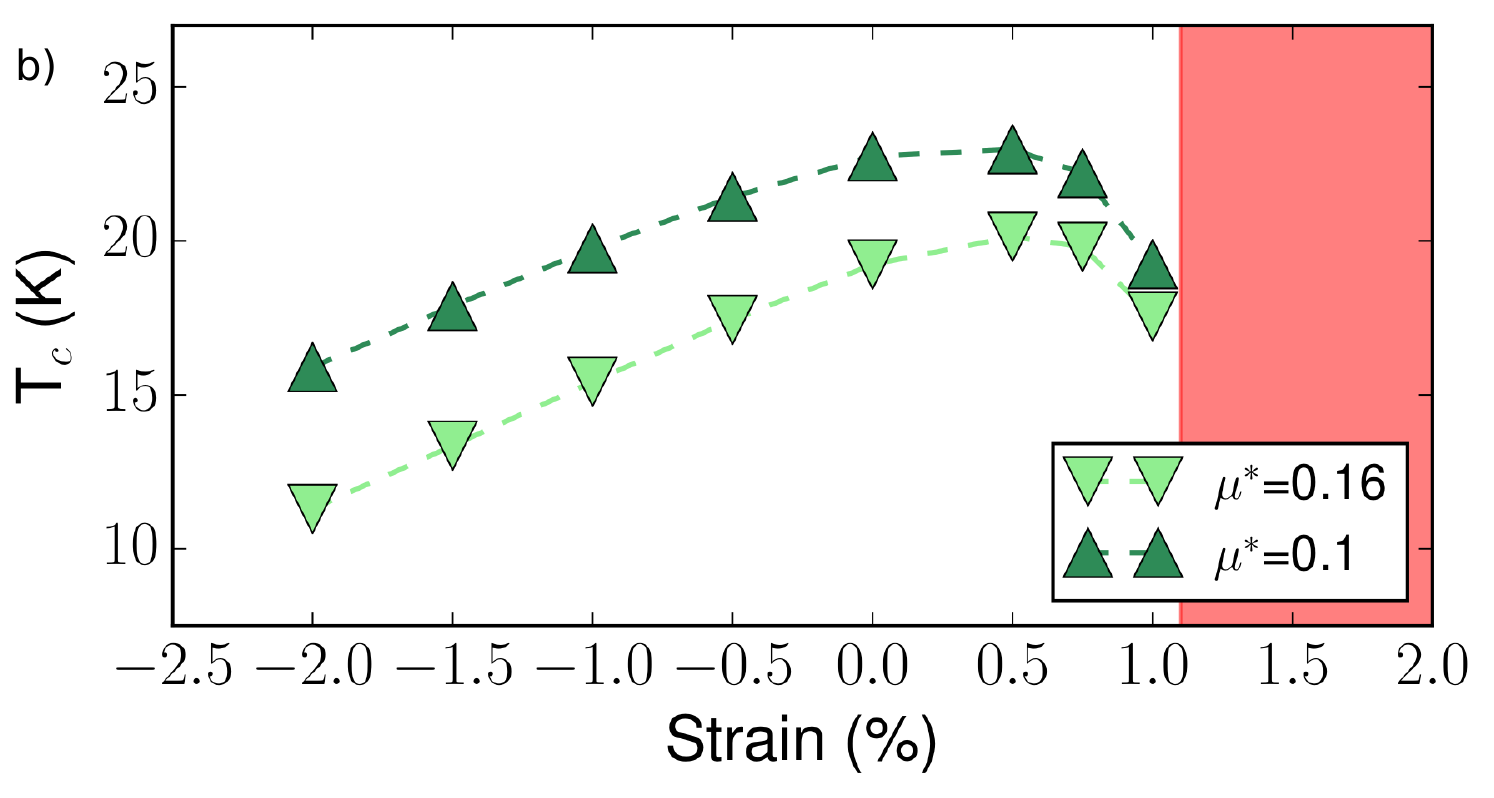}
\caption{a) Evolution of the electron-phonon coupling constant $\lambda$ and the logarithmically weighted frequency $\omega_{log}$ as a function of isostatic strain. b) Evolution of the superconducting transition temperature computed with the Allen-Dynes approximation and with two different standard values for the screened coulomb interaction $\mu^{\ast}$. }
\label{fig:straintc}
\end{figure}

\begin{figure}[h!]
\includegraphics[width=0.6\textwidth]{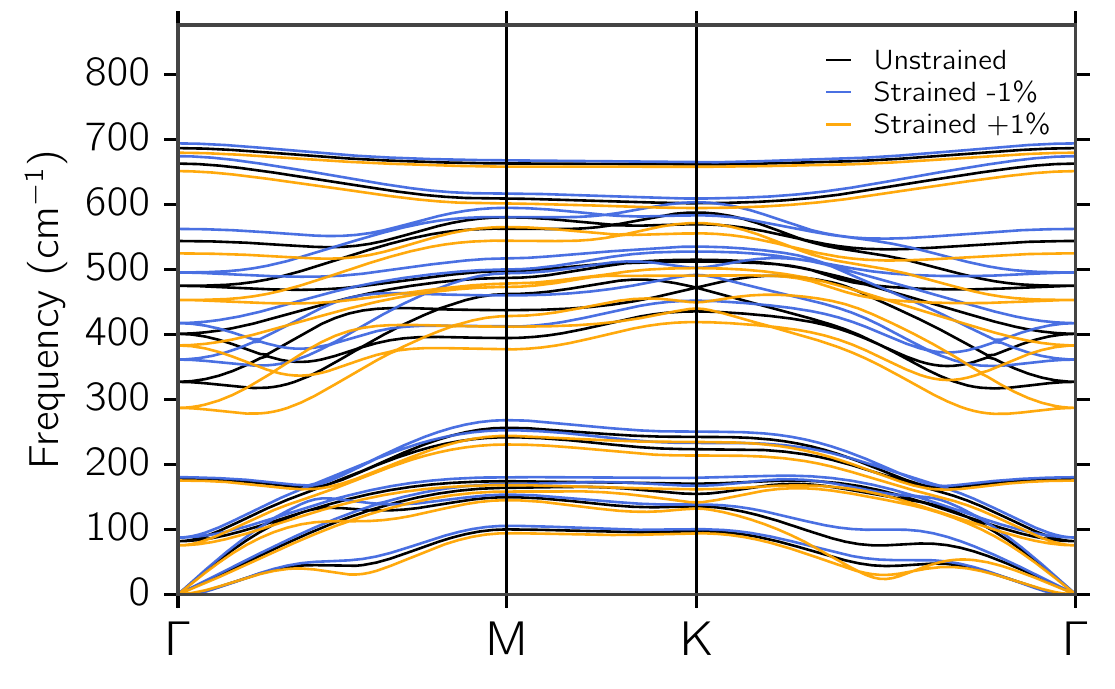}
\caption{Phonon dispersions along a standard high-symmetry path as a function of 1\% compressive (blue) and 1\% expansive (orange) isostatic strain compared with the unstrained case (black). The compression of the in-plane lattice parameter slightly stiffens the optical and the acoustic modes, while expansion softens both the optical modes and, more dramatically, the acoustic modes near the middle of the Brillouin zone, producing a more distinct anomalous kink in the phonon dispersions pointing to the incipient instability.}
\label{fig:phstrain}
\end{figure}

Strains up to few \% can be easily achieved in two dimensional materials by mechanical manipulation or substrate effects and, in Fig. \ref{fig:straintc} we report the  effects of a compressive and expansive isostatic deformation of the lattice parameter on the integrated electron-phonon coupling constant $\lambda$, the logarithmic frequency $\omega_{log}$ and the superconducting transition temperature T$_c$ estimated 
with the Allen-Dynes formula and the direct integration method detailed in the previous paragraph. 
The contraction of the lattice parameter brings a relatively slow decrease in the electron-phonon coupling constant that goes from $\lambda$=1.9 in the unstrained case down to $\lambda$=1.0 with a 2\% compression. The decrease in $\lambda$ is accompanied by an increase in $\omega_{log}$, signaling a diminishing role of the acoustic modes against a fairly constant coupling and blue shift of the optical modes. This compensation effect is reflected in the behavior of the superconducting transition temperature that shows a weaker decline from 19 to 12 K in the same strain range. On the other hand, even moderate expansions of the lattice parameter rapidly show dramatic effects on the electron-phonon couplings with $\lambda$ jumping to a value of 3.5 with at 1\% expansion. Such value would represent an absolute record, considerably higher than the highest electron-phonon coupling ever observed in nature ($\lambda$=2.6 in bismuth-lead alloys), and the experimental study of such extreme regime would very precious to understand the limitation of our current theories of superconductivity.
Unsurprisingly for higher strains the system becomes unstable with a transformation into a charge-density-wave driven by the highly interacting acoustic modes. A direct example of the effect of the strain on the phonon dispersions can be observed in Fig. \ref{fig:phstrain} highlighting how an expansion of 1\% in the lattice parameter induces a strong anomalous softening of the acoustic modes. The details on the effects of strain on the structure, phonon dispersions and electron-phonon couplings can be found in the S.I. 

\begin{figure}[h!]
\includegraphics[width=0.6\textwidth]{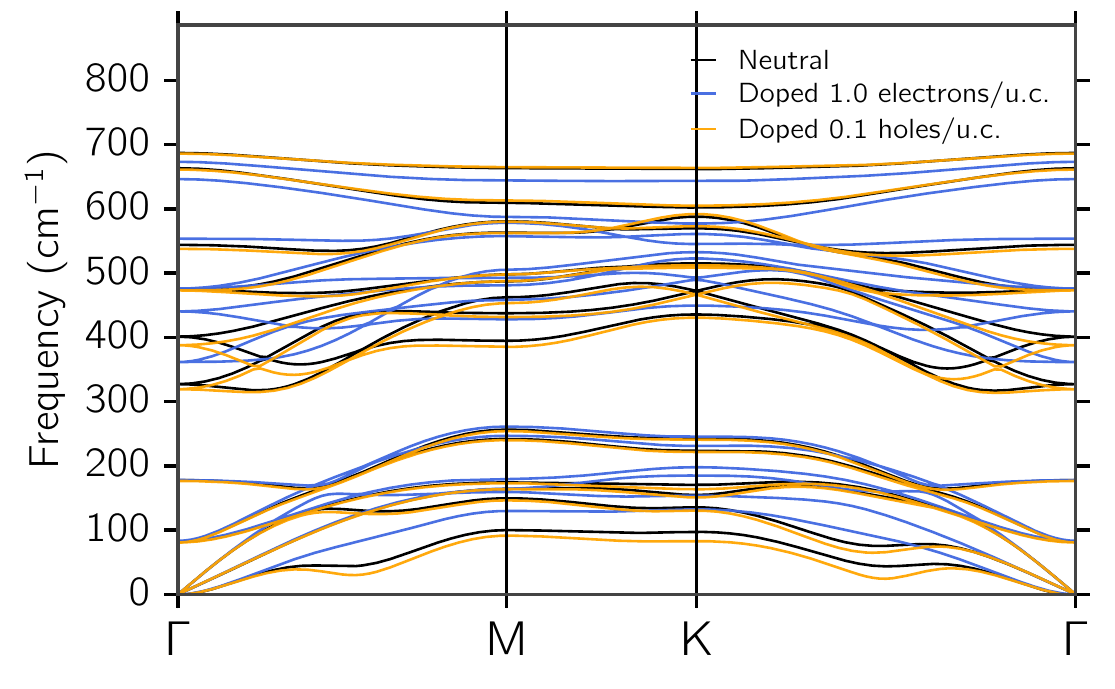}
\caption{Phonon dispersions along high-symmetry paths as a function of electron (negative) and hole (positive) doping. }
\label{fig:phdoping}
\end{figure}

\begin{figure}[h!]
\includegraphics[width=0.5\textwidth]{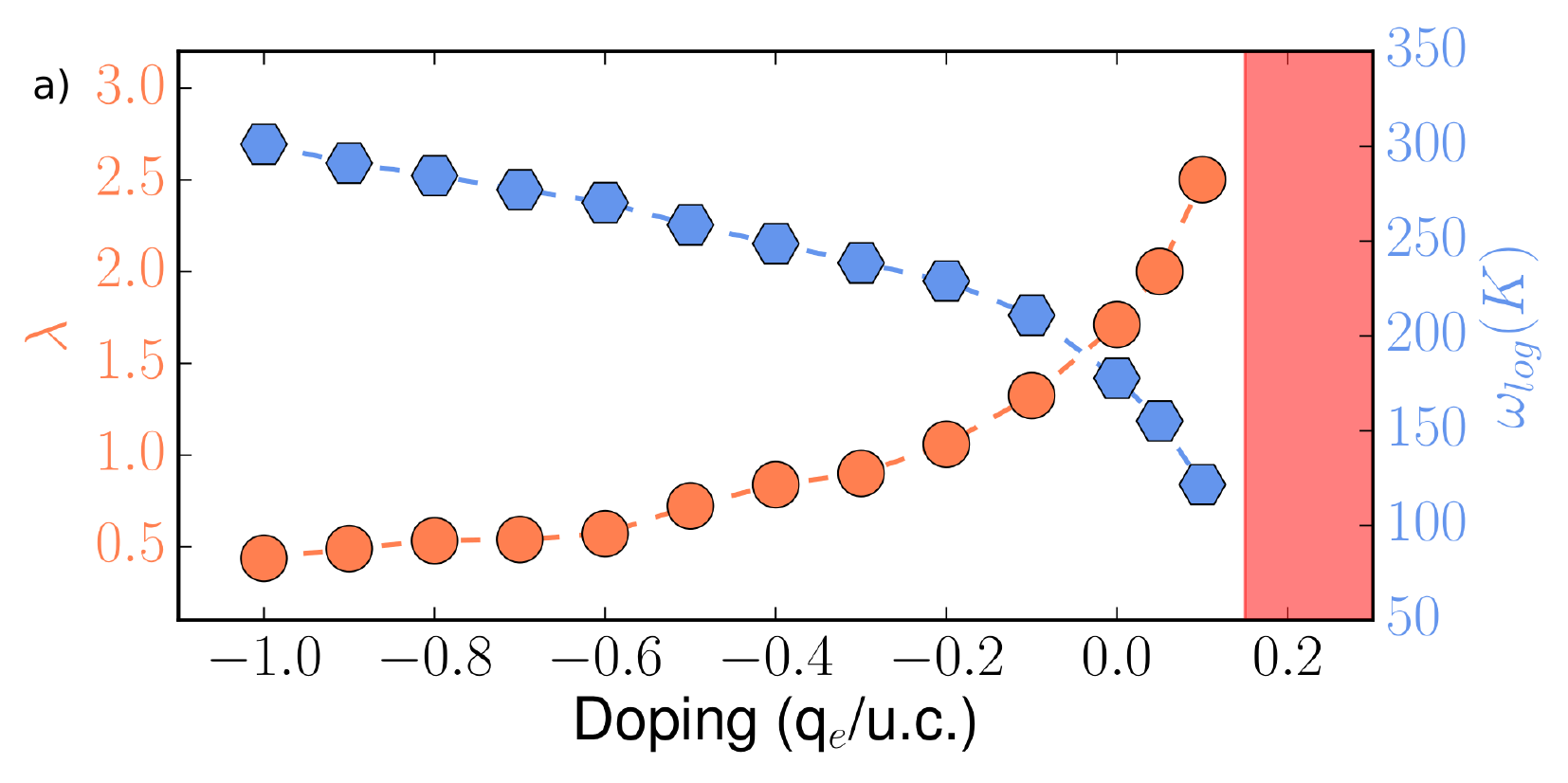}
\includegraphics[width=0.45\textwidth]{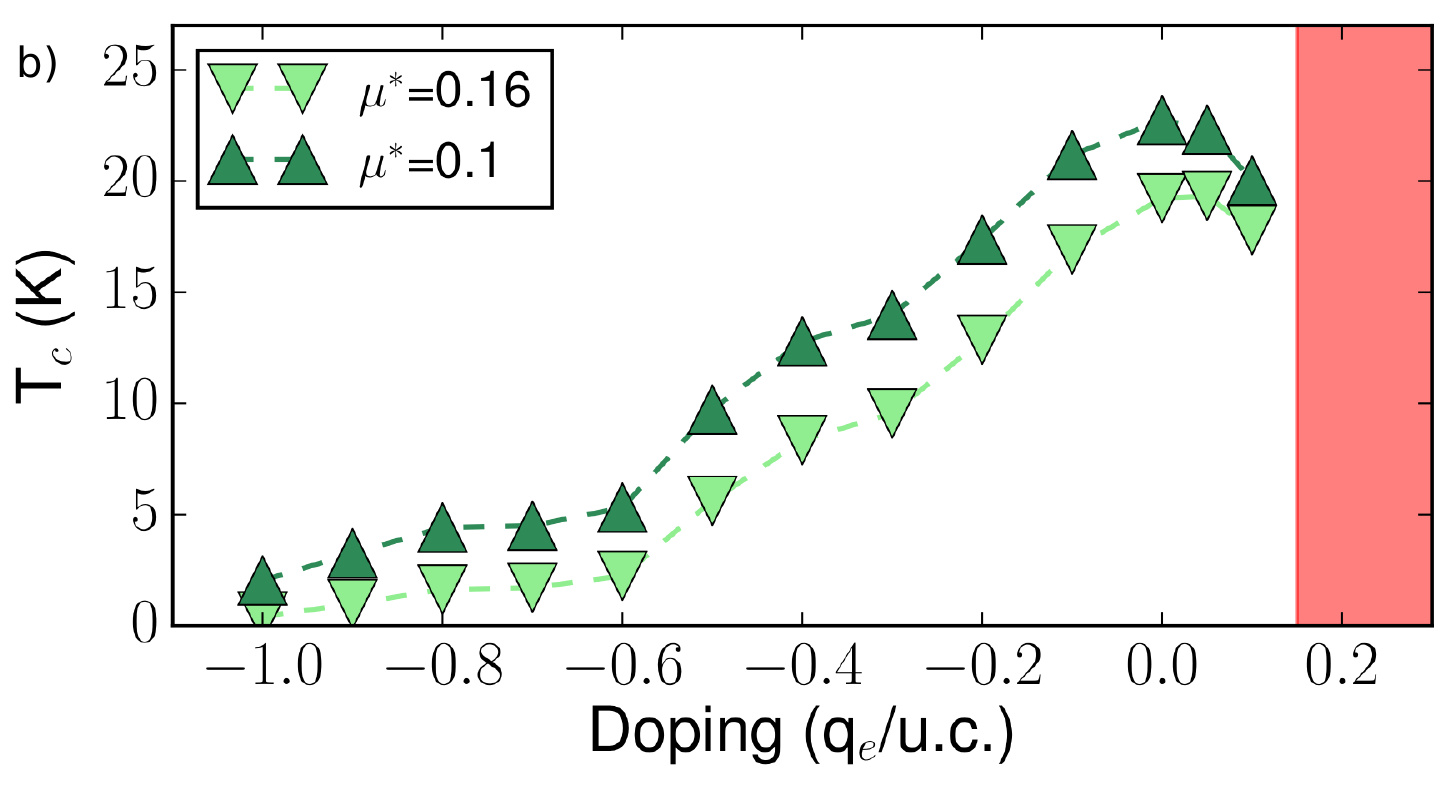}
\caption{a) Evolution of the electron-phonon coupling constant $\lambda$ and the logarithmically weighted frequency $\omega_{log}$ as a function of doping. b) Evolution of the superconducting transition temperature computed with the Allen-Dynes approximation using two different typical values for the screened Coulomb interaction $\mu^{\ast}$ }
\label{fig:dopingtc}
\end{figure}

In W$_2$N$_3$  heavy electron doping would be crucial to potentially exploit its topological properties, since the topologically protected bands appear far above the Fermi energy of the undoped material. We investigate the effects of doping  by introducing additional fractions of electron in the unit cell, compensated by a uniform background jellium, recomputing in each case the phonons and the phonon-induced potential response. This approach has been benchmarked by comparing it with a calculation in which actual lithium atoms absorbed on the surface (see S.I.). While hole doping rigidly shifts the band structure, electron doping has a more dramatic and non-trivial effect on the band structure, leading to the formation of free-electron-like bands, with parabolic dispersion centered at the $\Gamma$-point. These bands are reminiscent of the interlayer states observed in alkali-intercalated graphite\cite{Posternak1983,Holzwarth1984,Kaneko2017}. These bands don't give any substantial contribution to the charge density within the material; as such they don't interact significantly with the phonons of the monolayer and thus their presence  does not alter the nature of the electron-phonon couplings, that is still very much dominated by the hexagonally-warped states around the zone center. In the same way they don't interfere significantly with the presence of the edge states that should persist despite a likely overall break of the mirror symmetry induced by the most common doping methods. 
The effects of doping are summarized in Figs. \ref{fig:phdoping} and \ref{fig:dopingtc}, a moderate hole doping sensibly increases the electron-phonon coupling, leading to a softening of the acoustic modes similar to the one observed for expansive strain (Fig.\ref{fig:phdoping}) ultimately leading to a structural instability at a doping level above 0.175 holes per unit cell.

 At the very high electron doping regimes (0.8-1.0 electron per unit cell) achievable by alkali-metals intercalation and necessary to fill the topologically protected bands, the superconducting transition temperature deteriorates significantly, due to the dramatic decrease in the electron-phonon coupling constant as a consequence of the reduced nesting size in the central bands. The transition temperatures in this regime depend dramatically on the doping and the value of the screened coulomb interaction $\mu^{\ast}$ ranging from 0.4 to 1.6 K for $\mu^{\ast}$=0.16 and from 2 to 4.4 K for $\mu^{\ast}$=0.1. These values are in any case comparable or in some case significantly higher than the transition temperature for the topologically non trivial WTe$_2$ (0.9 K), for which  W$_2$N$_3$ could represent a valuable alternative.
 
Furthermore,  the strong dependence that electron-phonon couplings show with respect to both doping and strain in 2D W$_2$N$_3$ makes this material a unique platform to study different coupling regimes and the interplay between superconductivity and charge ordering derived from dynamical instability. 

\subsection{Conclusions}
In summary, we have discussed the topological and superconducting properties of W$_2$N$_3$ monolayer recently identified as easily exfoliable\citep{Mounet2018} from its 3D parent compound  and unveiled its rich physics. We predict a record-high transition temperature for a conventional phonon-mediated 2D superconductor of 21 K with fully anisotropic solution the Migdal-Eliashberg equations. We also highlight the effects of biaxial strain on the electron-phonon couplings and predict the marked dependence of the electron-phonon coupling constant, that makes 2D W$_2$N$_3$ a very promising platform to study different interaction regimes and test the limits of current theories of superconductivity. Finally, we discuss its topologically non-trivial bands resulting in unoccupied helical edge states 0.5 eV above the Fermi level;  the material could be  doped to fill such states and  superconductivity would persists, even if with a much reduced transition temperature,  making W$_2$N$_3$ also a viable candidate in the quest of exotic state of matter.

\acknowledgements

This work was supported by the Center for Computational Design and Discovery on Novel Materials NCCR MARVEL of the Swiss National Science Foundation.  D.C. also acknowledges the support from the ‘EPFL Fellows’ fellowship programme co-funded by Marie Sklodowska-Curie, Horizon 2020 grant agreement no. 665667. S.K. acknowledges the support of a MARVEL INSPIRE fellowship. 
Simulation time was awarded by CSCS on Piz Daint (production project s825) and by PRACE on  Marconi at Cineca, Italy (project id.\ 2016163963).

\clearpage

\bibliography{arxiv}

\end{document}